\documentclass[runningheads,tikz,14pt,border=10pt]{llncs}

\usepackage{amsthm,paralist}
\usepackage{mdframed}
\usepackage{centernot}
\usepackage{comment}
\usepackage{mathtools,xparse}
\usepackage{pifont}
\usepackage{amsfonts}

\usepackage{wrapfig}
\usepackage{wrapfig}
\usepackage{booktabs}
\usepackage{subcaption} 

\usepackage[ruled,vlined,linesnumbered,lined]{algorithm2e}

\usepackage{algpseudocode}
\usepackage{tikz}

\newcommand{\Ff}{\mathcal{F}}
\newcommand{\Bb}{\mathcal{B}}

\newcommand{\Nn}{\mathcal{N}}
\newcommand{\Ll}{\mathcal{L}}
\newcommand{\Pp}{\mathcal{P}}
\newcommand{\sPp}{\sem{\Pp}}

\newcommand{\Ss}{\mathcal{S}}

\newcommand{\sem}[1]{ [ \! [ {#1}  ]  \! ]} 
\def\rmdef{\stackrel{\mbox{\rm {\tiny def}}}{=}} 

\newcommand\vx{\mathbf{x}}

\newcommand\vd[2]{d_{i, p}}
\newcommand\vy{\mathbf{y}}

\newcommand{\set}[1]{\left\{ #1 \right\}}

\newcommand{\seq}[1]{\langle #1 \rangle}
\newcommand{\Nat}{\mathbb N}

\newcommand{\R}{\mathbb R}

\newcommand{\Real}{\R}
\newcommand{\Bool}{\mathbb B}
\newcommand{\Rplus}{\R_{\geq 0}}
\newcommand{\SE}{\textsf{SE}}

\theoremstyle{definition}

\numberwithin{exmp}{section}

\definecolor{gold}{rgb}{0.99,0.78,0.07}
\usetikzlibrary{arrows,shapes,snakes,automata,backgrounds,positioning,decorations.pathmorphing,
	decorations.markings,calc}

\tikzstyle{dtreenode}=[draw=blue!10!gray,rounded rectangle, minimum size=5mm,fill=blue!10!white]
\tikzstyle{dtreeleaf}=[draw=black!60,minimum width=1cm,minimum height=0.4cm,rectangle,fill=blue!50!white]
\tikzset{every loop/.style={looseness=7}}
\tikzset{
	gluon/.style={decorate,draw=black,
		decoration={coil,amplitude=1pt, segment length=5pt}}
}
\tikzset{
	gluon1/.style={decorate,draw=black,
		decoration={coil,amplitude=3pt, segment length=3pt}}
}
\tikzset{
	gluonew/.style={decorate,draw=black,
		decoration={coil,amplitude=1pt, segment length=2pt}}
}

\tikzset{bicolor/.style args={#1 and #2 and #3}{
		path picture={
			\tikzset{rounded corners=0}
			\fill [#1] (path picture bounding box.south west)
			rectangle
			($(path picture  bounding box.north west)!#3!(path picture bounding
			box.north east)$);
			\fill [#2]
			($(path picture bounding box.south west)!#3!(path picture bounding
			box.south east)$)
			rectangle (path picture bounding box.north east);
}}}

\tikzset{tricolor/.style args={#1 and #2 and #3 and #4 and #5}{
		path picture={
			\tikzset{rounded corners=0}
			\fill [#1] (path picture bounding box.south west)
			rectangle
			($(path picture  bounding box.north west)!#4!(path picture bounding
			box.north east)$);
			\fill [#2]
			($(path picture bounding box.south west)!#4!(path picture bounding
			box.south east)$)
			rectangle
			($(path picture  bounding box.north west)!#5!(path picture bounding
			box.north east)$);
			\fill [#3]
			($(path picture bounding box.south west)!#5!(path picture bounding
			box.south east)$)
			rectangle (path picture bounding box.north east);
}}}

\usepackage{listings}

 \definecolor{dkgreen}{rgb}{0,0.6,0}
 \definecolor{gray}{rgb}{0.5,0.5,0.5}
 \definecolor{mauve}{rgb}{0.58,0,0.82}

\definecolor{cadmiumgreen}{rgb}{0.0, 0.42, 0.24}
\definecolor{verde}{rgb}{0.25,0.5,0.35}
\definecolor{jpurple}{rgb}{0.5,0,0.35}
\definecolor{darkgreen}{rgb}{0.0, 0.2, 0.13}

\newsavebox{\mybox}

\PassOptionsToPackage{hyphens}{url}\usepackage{hyperref}

\tikzset{%
  every neuron/.style={
    circle,
    draw,
    minimum size=0.8cm
  },
  neuron missing/.style={
    draw=none,
    scale=2,
    text height=0.111cm,
    execute at begin node=\color{black}$\vdots$
  },
}

\usepackage{textcomp}
\usetikzlibrary{shapes,arrows}

\title{Efficient Detection and Quantification of Timing Leaks with
Neural Networks}
\titlerunning{Quantification of Leaks with NNs}
\author{
  Saeid Tizpaz-Niari
  \and
  Pavol {\v C}ern\'y
  \and \\
  Sriram Sankaranarayanan
  \and
  Ashutosh Trivedi
}
\institute{University of Colorado Boulder \\
  \email{\{saeid.tizpazniari,
    pavol.cerny, \\srirams, ashutosh.trivedi\}@colorado.edu}}
\authorrunning{Tizpaz-Niari, et al.}

\begin{document}

\maketitle

\begin{abstract}
Detection and quantification of information leaks through timing side
channels are important to guarantee confidentiality.
Although static analysis remains the prevalent approach for detecting
timing side channels, it is computationally challenging 
for real-world applications.
In addition, the detection techniques are usually restricted to ``yes'' or
``no'' answers.
In practice, real-world applications may need to leak information about the
secret.
Therefore, quantification techniques are necessary to evaluate the resulting
threats of information leaks. 
Since both problems are very difficult or impossible for static analysis
techniques, we propose a dynamic analysis method. 
Our novel approach is to split the problem into two tasks.
First, we learn a timing model of the program as a neural network.
Second, we analyze the neural network to quantify information leaks.
As demonstrated in our experiments, both of these tasks are feasible in practice
--- making the approach a significant improvement over the state-of-the-art side
channel detectors and quantifiers. 
Our key technical contributions are (a) a neural network architecture
that enables side channel discovery and (b) an MILP-based algorithm
to estimate the side-channel strength.
On a set of micro-benchmarks and real-world applications, we show that neural
network models learn timing behaviors of programs with thousands of methods.
We also show that neural networks with thousands of neurons can be efficiently
analyzed to detect and quantify information leaks through timing side channels.

\end{abstract}

\section{Introduction}
\label{sec:introduction}
Programs often handle sensitive data such as credit card numbers or medical
histories. Developers are careful that eavesdroppers cannot easily access the
secrets (for instance, by using encryption algorithms). However, a side
channel might arise even if the transferred data is encrypted. For example, in
timing side channels~\cite{brumley2005remote}, an eavesdropper who observes the
response time of a server might be able to infer the secret input, or at least
significantly reduce the remaining entropy of possible secret values.
Studies show that side-channel attacks are
practical~\cite{kocher1996timing,CWWZ10,hund2013practical}.

Detecting timing side channels are difficult problems
for static analysis, especially in real-world Java applications.
From the theoretical point of view, side-channel presence
cannot be inferred from one execution trace, but rather, an analysis of
equivalence classes of traces is needed~\cite{sabelfeld2003language}.
From a practical perspective,
the problem is hard because timing is not explicitly visible in the code.
A side channel is a property of the code and the platform on which
the program is executed.
In addition, most existing static techniques rely on taint
analysis that is computationally difficult for applications with
dynamic features~\cite{landman2017challenges}.

A large body of work addresses the problem of detecting
timing side channels~\cite{DBLP:conf/ccs/ChenFD17,DBLP:conf/icse/nilizadeh,antonopoulos2017decomposition}.
However, detection approaches
are often restricted to either ``yes'' or ``no'' answers.
In practice, real-world applications may need to
leak information about the secret~\cite{smith2009foundations}.
Then, it is important to know ``how much'' information is being
leaked to evaluate the resulting threats.
Quantification techniques with entropy-based
measures are the primary tools to calculate the amount of leaks.

We propose a data-driven dynamic analysis for detecting and quantifying
timing leaks.
Our approach is to split the problem into two tasks: first, to learn a timing
model of the program as the neural network (NN) and second, to analyze the NN,
which is a simpler object than the original program. The key insight
that we exploit is that {\em timing models of the program are
easier to learn than the full functionality of the program}. The advantages of
this approach are two-fold. First, although general verification problems are
difficult in theory and practice, learning the timing models is
efficiently feasible, as shown in our experiments. We conjecture that this is
because timing behaviors reflect the computational complexity of a program, which is
usually a simpler function of the input rather than the program. Second,
neural networks, especially with ReLU units, are easier models to analyze than
programs.

Our key technical ideas are enabling a side channel analysis using a specialized
neural network architecture and a mixed integer linear programming (MILP) algorithm
to estimate an entropy-based measure of side-channel strengths. First, our NN
consists of three parts: (a) encoding the secret inputs, (b) encoding the public
inputs, and (c) combining the outputs of the first two parts to produce the
program timing model. This architecture has the advantage that we can easily
change the ``strength" (the number of neurons $k$) of the connection from the secret
inputs. This enables us to determine whether there is a side channel.
If timing can not be accurately predicted based on just public inputs ($k {=} 0$), then
there is a timing side channel. Second, for $k {>} 0$, the trained NN
can be analyzed to estimate the strength of side-channel leaks.
Each valuation of the $k$ binarized output
of secret part corresponds to one
observationally distinguishable class of secret values.
We use an MILP encoding to estimate the size of classes and
calculate the amount of information leaks with entropy-based
objectives such as {\em Shannon} entropy.

Our empirical evaluation shows both that timing models of programs can be
represented as neural networks and that these NNs can be analyzed
to discover side channels and their strengths. We implemented our techniques
using Tensorflow~\cite{abadi2016tensorflow} and Gurobi~\cite{gurobi}
for learning timing models as NN objects and analyzing the NNs
with MILP algorithms, respectively.
We ran experiments on a Linux machine with 24 cores of 2.5 GHz.
We could learn timing model of programs (relevant to the observed traces)
with few thousands methods in less than 10 minutes with the accuracy
of 0.985.
Our analysis can handle NN models with thousands of neurons
and retrieve the number of solutions for each class of observation.
The number of solutions enables us to quantify the amount of information leaks.

\noindent In summary, our key contributions are:
\begin{compactitem}
  \item An approach for side channel analysis based on learning execution
  times of a program as a NN.
  \item A tunable specialized NN architecture for timing side-channel analysis.
  \item An MILP-based algorithm for estimating entropy-based measures of information
  leaks over the NN.
  \item An empirical evaluation that shows our approach is scalable and quantifies
  leaks for real-world Java applications.
\end{compactitem}

\section{Overview}	
\label{sec:overview}
We illustrate on examples the key ingredients of our approach: learning timing
models of programs using neural networks and a NN architecture that
enables us to detect and quantify timing leaks.

\begin{figure*}[t!]
   \begin{subfigure}{0.45\textwidth}
   \includegraphics[width=1.0\textwidth]{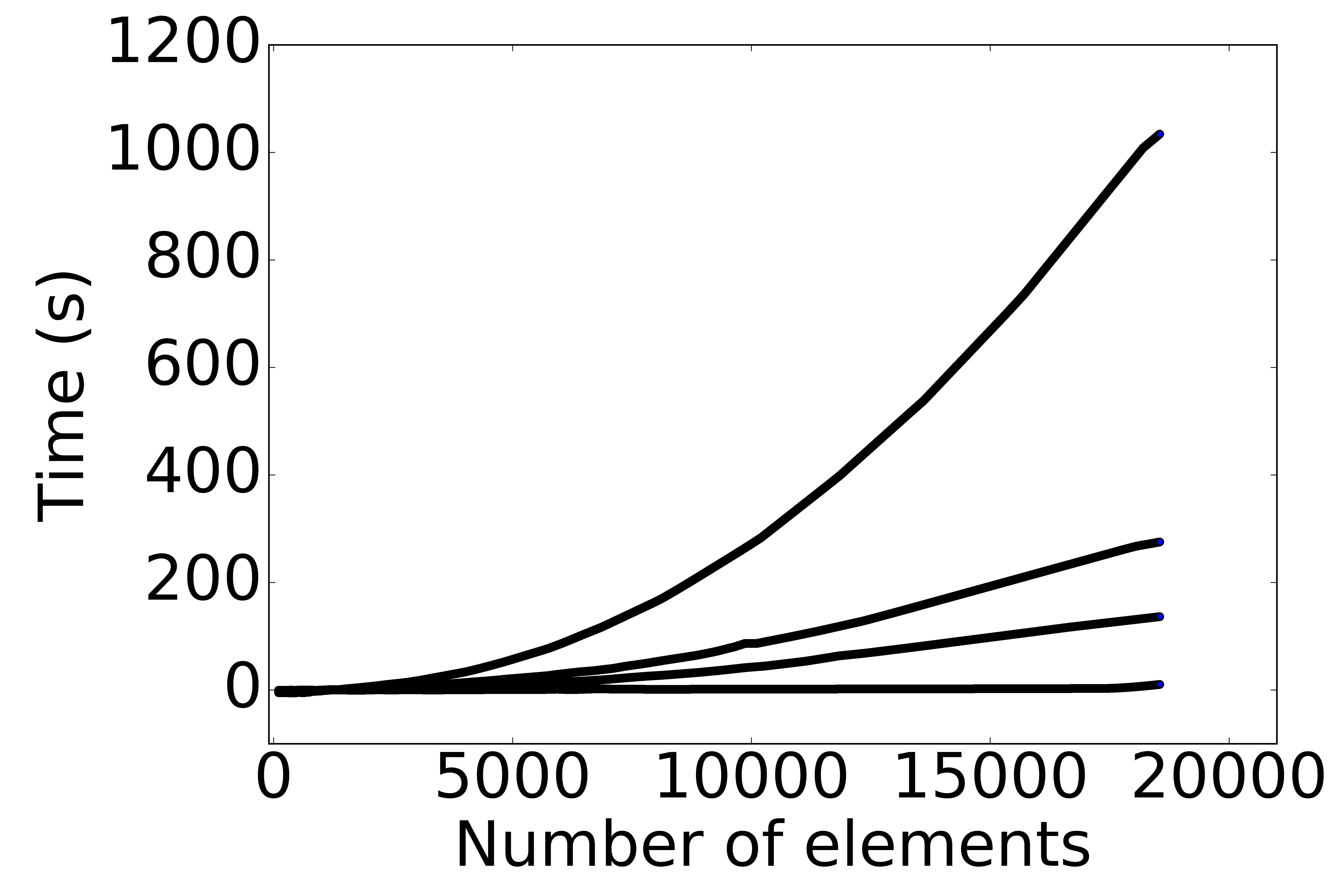}
   \end{subfigure}
   \hfill
   \begin{subfigure}{0.45\textwidth}
     \includegraphics[width=1.0\textwidth]{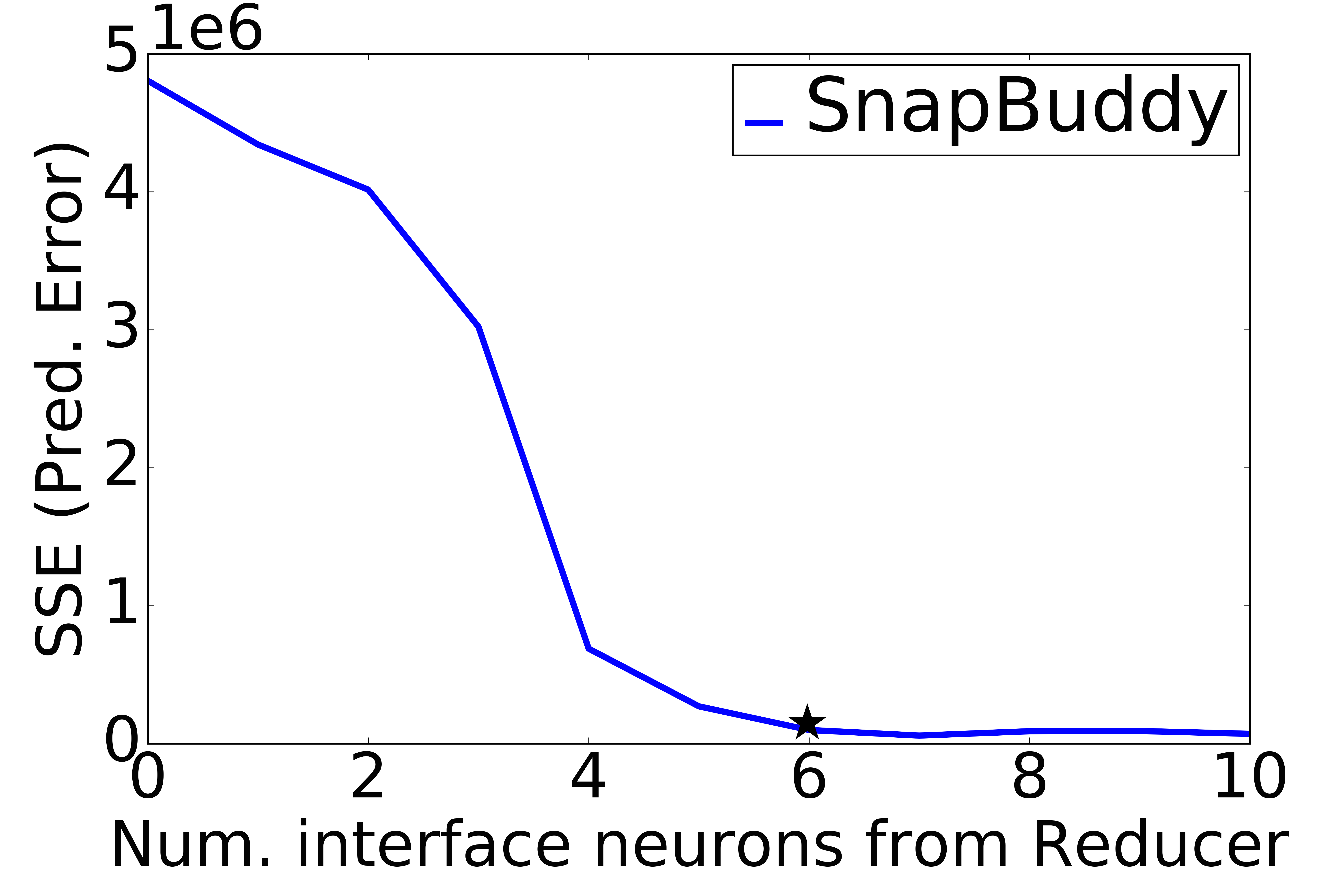}

   \end{subfigure}
   \caption{(a) Timing models of the sorting application as learned by NNs.
   (b) Prediction error vs. the number of neurons ($k$) in
   the interface layer of SnapBuddy.}
   \label{fig:SB-SSE-k}
   \label{fig:sort}
   \vspace{-1.0em}
 \end{figure*}

\noindent{\bf Learning timing models of programs.}
 Learning the functionality of the sorting algorithms from programs is
 difficult, while learning their timing model is much easier. We picked the
 domain of sorting algorithms as these algorithms have well-known different
 timing behaviors.
 We implement six different sorting algorithms namely bubble sort, selection
 sort, insertion sort, bucketing sort, merging sort, and quick sort in one
 sorting application. We generate random arrays of different sizes from
 100 to 20,000 elements and
 run the application for each input array with different sorting algorithms.
 This implies that we consider the average-case
 computational complexities of different algorithms. Each data point
 consists of the input array, the indicator of sorting algorithm,
 and the execution time of sorting application.
 In addition, the data points are independent from each other, and
 the neural network learns end-to-end execution times of the
 entire application, not individual sorting algorithms.
 Figure~\ref{fig:sort} (a) shows the learned
 execution times for the sorting application. The neural
 network model has 6 layers with 825 neurons. The
 accuracy of learning based on coefficient of determination ($R^2$) over
 test data is 0.999, and the learning takes 308.7 seconds
 (the learning rate is 0.01). As we discussed earlier, the NN model
 predicts (approximates) execution times for
 the whole sorting application. Further analysis is required to
 decompose the NN model for individual sorting algorithms that is
 not relevant in our setting since we are only interested in learning
 end-to-end execution times of applications.
 Using the neural network model to estimate
 execution times has two important advantages over the state-of-the-art
 techniques~\cite{goldsmith2007measuring,aaai18}. First, the neural network can
 approximate an arbitrary timing model as a function of input features, while
 the previous techniques can only approximate linear or polynomial functions.
 Second, the neural network does not require feature engineering, whereas the
 previous techniques require users to specify important input features such
 as work-load (size) features.

\begin{figure}[t]
\begin{center}

\begin{tikzpicture}[auto, thick, node distance=2cm, >=triangle 45]

\tikzset{%
  operator/.style = {draw,fill=yellow!10,minimum size=1.2em},
  operator2/.style = {draw,fill=blue!10,minimum height=1.6cm},
  crossx/.style={path picture={
      \draw[thick,black,inner sep=0pt]
      (path picture bounding box.south west) -- (path picture bounding
      box.north east);
      }},
}


\draw
	node [name=input1] at (-1.9,0.0)[right=-1mm]{${\vx \in \Real^n}$}
  node [name=input2] at (-1.9,-2.0)[right=-1mm]{${\vy \in \Real^m}$}
	node [operator2, right of=input1] (layer1S) {$\Ll_1^S$}
	node at (2.0,0.0) (dots) {\ldots}
  node at (3.2,0.0) [operator2] (layerNS) {$\Ll_N^S$}
  node at (4.7,0.0) [operator] (layerAlpha) {$\Ll_\alpha^S$}
  node at (6.2,0.0) [circle,fill=black!10] (boolK){$\Bool^k$}
  node [operator2, right of=input2] (layer1P) {$\Ll_1^P$}
  node at (2.0,-2.0) (dots2) {\ldots}
  node at (3.5,-2.0) [operator2] (layerNP) {$\Ll_M^P$}
  node at (6.3,-1.3) [operator2] (layer1J) {$\Ll_1^J$}
  node at (7.7,-1.3) (dots3) {\ldots}
  node at (9.0,-1.3) [operator2] (layerNJ) {$\Ll_R^J$}
  node at (10.1,-1.3) (output) {${\hat{t}}$}
;
	\draw[->](input1) -- node {\tiny $~~W^{1}_{s}$}(layer1S);
 	\draw[->](layer1S) -- node {\tiny $W^{2}_{s}~~$} (dots);
	\draw[->](dots) -- node {} (layerNS);
  \draw[->](layerNS) -- node {\tiny $W^{N+1}_{s}$} (layerAlpha);
  \draw[->](input2) -- node {\tiny $W^{1}_{p}$}(layer1P);
  \draw[->](layer1P) -- node {\tiny $W^{2}_{p}~~$} (dots2);
  \draw[->](dots2) -- node {} (layerNP);
  \draw[->](layerAlpha) -- node {} (boolK);
  \draw[->](layerAlpha) -- node[pos=0.1,rotate=-40] {\tiny $W^{1}_{s \to j}$} (layer1J);
  \draw[->](layerNP) -- node[rotate=10] {\tiny $~~W^{1}_{p \to j}$} (layer1J);
  \draw[->](layer1J) -- node {\tiny $W^{2}_{j}~~$} (dots3);
  \draw[->](dots3) -- node {} (layerNJ);
  \draw[->](layerNJ) -- node {} (output);

	\draw [color=gray,line width=0.7mm](-0.5,-3) rectangle (9.5,1);
  \draw [color=blue!80,line width=0.5mm](-0.4,-0.9) rectangle (5.1,0.9);
  \node at (4.98,0.8) [color=blue!80,above=5mm, right=0mm] {\textsc{Reducer Function:}$\Real^n \to \Bool^k$};
\end{tikzpicture}
\end{center}
\caption{Neural network architecture for side-channel discovery
and quantification. The NN takes secret features ($\vx$) and public
features ($\vy$), and it learns weights $W$ with the minimal
number of neurons in the interface layer ($\Ll_\alpha^S$)
to precisely predict execution times.
The reducer function maps
$n$-dimensional $\vx$ to $k$ binary outputs. The MILP
analysis of reducer function is used for the quantification.}
\label{fig:arch}
\end{figure}
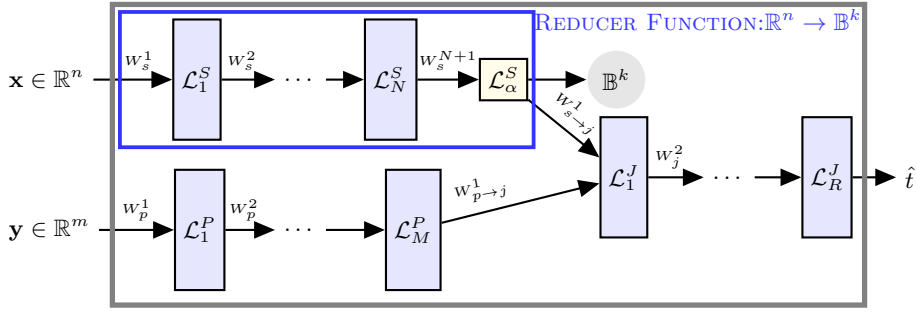

\noindent{\bf Neural network architecture for side channel discovery.}
With the observations that the neural networks can learn programs' execution
times precisely, we consider a special architecture to analyze timing side
channels of programs. We propose the neural network architecture shown in
Figure~\ref{fig:arch}. The NN architecture consists of three parts: 1) A
reducer function that learns a map from $n$ secret features (inputs) to $k$
binarized  interface neurons (we call them interface neurons as they connect the
secret inputs to the rest of the network); 2) A neural network function
that connects the public
features to the overall model; 3) A joint function that uses the output of the
reducer and public-features functions to predict the execution times.
The architecture
makes it easy to change the number of neurons ($k$)
in the connection from the reducer, which enables us to estimate
the side-channel strength.
In this architecture, there are timing side channels if the value for
$k$ is greater than or equal to 1.
The NN learning is to find the weights in different layers
with the optimal number of neurons in the interface layer ($k$) such that
the NN approximates execution times accurately.

\noindent{\bf Estimating the side channel strength.}
The side-channel strength is estimated by finding the minimal value of $k$
in the learning of accurate NN models.
We use Sum of Squared Error (SSE) measure to compute prediction errors.
Figure~\ref{fig:SB-SSE-k} (b) shows the SSE versus the number of interface
neurons ($k$) for the SnapBuddy application (described in
Section~\ref{sec:case-SB}).
As shown in the plot, the prediction error
decreases as the number of neurons increase from 0 to 6. But, after 6,
the prediction error stays almost the same. We thus choose 6
as the optimal number of interface neurons.
Since each neuron is a binary unit, there
are $2^6$ distinct outputs from the reducer function.
Each distinct output
forms a class of observation
over the secret inputs. However, some classes of observations
might be empty and are not
feasible from any secret value. Furthermore,
for feasible classes, the entropy
measures require the number of elements in each class.
We encode the reducer function as a
mixed integer linear programming (MILP) problem. Then,
we calculate the number of feasible
solutions for each class.
For SnapBuddy, it takes 16.6 seconds to analyze the
reducer function and find the number of solutions for non-empty classes.
Using Shannon entropy, the analysis shows 3.0 bits of information
about secret inputs are leaking in SnapBuddy.

\section{Problem Statement}
\label{sec:definition}
We develop a framework for detecting and quantifying information leaks due to
the execution time of programs.
Our framework is suitable for
{\em known-message} and {\em chosen-message}
threat~\cite{kopf2009provably} settings where the variations in the
execution times depend on both public and secret inputs.

The {\it timing model } $\sPp$  of a program $\Pp$ is a tuple $(X,
Y, \Ss, \delta)$ where $X {=} \{x_1, \ldots$ $, x_n\}$ is the set of
{\it secret} input variables, $Y = \set{y_1, y_2, \ldots, y_m}$ is the set of
{\it public} input variables, and $\Ss \subseteq \Real^n$ is a finite set of {\it
secret} inputs, and  $\delta: \Real^n \times \Real^m \to \Rplus$ is the
execution-time of the program as a function of secret and public inputs.
A {\it timing function} of the program $\Pp$ for a secret
input $s \in \Ss$ is the function  $\delta(s)$ defined as
$\vy \in \Real^m \mapsto \delta(s, \vy)$.
Let $\Ff$ be the set of all timing functions in $\Pp$.

Given a timing model $\sPp$ and a tolerance $\varepsilon {>} 0$, a $k$-bit
$\varepsilon$-approximate secret reducer is a pair
\[
(\alpha, \beta) \in [\Real^n \to \Bool^k] \times[\Bool^k \to \Real^n]
\]
such that
$\|\delta(\vx) - \delta(\beta(\alpha(\vx)))\| \leq \varepsilon$
for every $\vx \in \Ss$ where $\|\cdot\|$ is some fixed norm over the space of
timing function. In this paper, we work with $\infty$-norm.
We write $R_{(\varepsilon, k)}$ for the set of all $k$-bit
$\varepsilon$-approximate reducers for $\sPp$.
We say that $(\alpha, \beta) \in R_{(\varepsilon, k)}$ is an optimal
$\varepsilon$-approximate reducer if for all $k' < k$ the set
$R_{(\varepsilon, k')}$ is empty.
Given a tolerance $\varepsilon > 0$, we say that there are information leaks
in execution times, if there is no $0$-bit $\varepsilon$-approximate optimal
secret reducer.

A reducer $(\alpha, \beta)$ characterizes an equivalence relation
$\equiv_{\alpha}$ over the set of secrets $\Ss$, defined as the following:
$s \equiv_{\alpha} s'$ if $\alpha(s) = \alpha(s')$.
Let $\Ss_{[\alpha]} = \seq{\Ss_1, \Ss_2, \ldots, \Ss_K}$ be the quotient space of $\Ss$
characterized by the reducers $(\alpha, \beta)$; note that $2^{k-1} < K \leq 2^{k}$.
Let $\Bb = \seq{B_1, B_2, \ldots, B_{K}}$ be the size of observational
equivalence class in $\Ss_\alpha$, i.e. $B_i = |\Ss_i|$ and let $B =
|\Ss| = \sum_{i=1}^K B_i$.
The expected information leaks due to observations on the execution
times of a program
can be quantified by using the difference between the uncertainty about the secret
values before and after the timing observations.
Assuming that secret values $\Ss$ are uniformly distributed, we quantify
information leaks~\cite{KB07} as
\begin{equation}
\label{eq:se}
\SE(\Ss|\alpha) \rmdef {\log_2(B)} - \frac{1}{B} \sum\limits_{i{=}1}^{K} B_i
\log_2(B_i).
\end{equation}
Given a program with inputs partitioned into secret and public inputs, our
goal is to quantify the information leaks through timing side channels.
However, such programs often have complex functionality with black-box
components. Moreover, the shape of timing functions may be non-linear and
unknown.
We propose a neural-network architecture to approximate the timing model as
well as to quantify information leakage due to the timing side channels
in the program.
We then analyze this network to precisely quantify the information leaks
based on the equation (\ref{eq:se}).

\section{Neural Network Architecture to Detect and Quantify Information Leaks}
\label{sec:neural}
A rectified linear unit  (ReLU) is a function $\sigma: \Real \to
\Real$ defined as $x \mapsto \max \set{x, 0}$.
We can generalize this function from scalars to vectors as $\sigma: \Real^n \to
\Real^n$  in a straightforward fashion by applying ReLU component-wise.
In this paper, we primarily work with feedforward neural network (NN) with ReLU
activation units.
A $\Real^{w_0} \to \Real^{w_{N+1}}$ feedforward neural network $\Nn$ is characterized by its
number of hidden layers (or depth) $N$, the input and output dimensions $w_0,
w_{N+1} \in \Nat$, and width of its hidden layers $w_1, w_2, \ldots, w_N$.
Each hidden layer $i$ implements an affine mapping $T_i :\Real^{w_{i-1}} \to
\Real^{w_i}$ corresponding to the weights in each layer.
The function $f_\Nn: \Real^{w_0} \to \Real^{w_{N+1}}$ implemented by
neural network $\Nn$ is:
\[
f_\Nn = T_{k+1} \circ \sigma \circ T_k \circ \sigma \circ \cdots T_2 \circ \sigma
\circ T_1.
\]
It is well known that NNs with ReLU units implement a piecewise-linear
function~\cite{ABMM16} and due to this property, it can
readily be encoded~\cite{Fischetti2018} as a mixed integer linear
programming (MILP).

Given a target (black-box) function  $f: \Real^{w_0} \to \Real^{w_{N+1}}$ to be
approximated and a neural network architecture $(w_0, w_1, \ldots, w_{N+1}) \in
\Real^{N+2}$,  the process of training the network is to search for weights
of various layers so as to closely approximate the function $f$
based on noisy approximate examples from the function $f$.
The celebrated universal approximation theorems about neural networks state that
\emph{deep feedforward neural networks}~\cite{Cyb89,Hornik89}---equipped with
simple activation units such as rectified linear unit (ReLU)---can approximate
arbitrary continuous functions on a compact domain to an arbitrary precision.
Assuming that the timing functions of a program have bounded
discontinuities, it can be approximated with a continuous function to an
arbitrary precision. It then follows that one can  approximate the
execution-time function to an arbitrary precision using feedforward neural
networks.

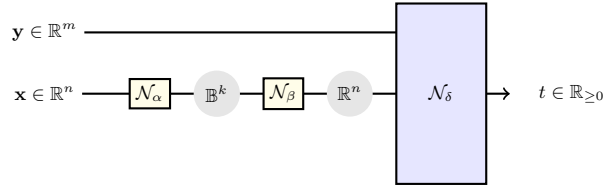
\begin{figure}[t]
\begin{center}
\begin{tikzpicture}[thick, scale=0.8, every node/.style={scale=0.8}]
  
\tikzset{
  operator/.style = {draw,fill=yellow!10,minimum size=1.5em},
  operator2/.style = {draw,fill=blue!10,minimum height=3cm},
  crossx/.style={path picture={
      \draw[thick,black,inner sep=0pt]
      (path picture bounding box.south west) -- (path picture bounding
      box.north east);
      }},
}
\matrix[row sep=-0.6cm, column sep=0.3cm] (circuit) { 
  \node (q1) {$\vy \in \Real^m$};
  &   &  &  &  &  &  \coordinate (end1); 
  &
  &\\
  \node (q2) {$\vx \in \Real^n$};
  &
  &\node[operator] (H21) {$\Nn_\alpha$};
  &\node[circle,fill=black!10] (c21){$\Bool^k$};
  &\node[operator] (H21) {$\Nn_\beta$};
  &\node[circle,fill=black!10] (c22){$\Real^n$};
  &\node[operator2](U21){$~~~~\Nn_\delta~~~~$};
  \coordinate (end2);
  &\coordinate (end3); 
  &\node(q3) {$t \in \Rplus$};\\
};

\begin{pgfonlayer}{background}
  \draw[thick,->] (q1) -- (end1)
  (q2) -- (end2)
  (end2) -- (end3);
\end{pgfonlayer}
\end{tikzpicture}

\end{center}
\caption{Neural Network $\Nn_\delta$ approximating the execution-time function
  $\delta$ along with a reducer $(\Nn_\alpha, \Nn_\beta)$.}
\label{fig:nn-reducer}
\end{figure}

Figure~\ref{fig:nn-reducer} shows different components of our
neural network model.
We train a neural network $\Nn_\delta: \Real^{n+m} \to \Real$ (where the input
variables are partitioned into secret and public and the output variable is
the execution time) to approximate the execution times of a given program to a
given precision $\varepsilon >0$.
In order to quantify the number of secret bits leaked in the timing
functions, we train a pair of $k$-bit reducer neural
networks $\Nn_\alpha: \Real^n \to \Bool^k$ and $\Nn_\beta: \Bool^k \to \Real^n$
with the output of $\Nn_\beta$ connected to the neural network $\Nn_\delta$.
In this training, we only learn the weights of $\Nn_\alpha$ and $\Nn_\beta$
while keeping the weights of $\Nn_\delta$ unchanged.
We call the composition of these networks
$\Nn_k = \Nn_\alpha\circ\Nn_\beta\circ\Nn_\delta$.
It is easy to see that the network pair $(\Nn_\alpha, \Nn_\beta)$ implements a
$k$-bit secret reducer $(f_{\Nn_\alpha}, f_{\Nn_\beta})$.
Let $k$ be the smallest number such that the fitness of $\Nn_k$ is comparable to
the fitness of $\Nn_\delta$.
We find the smallest $k \in \Nat$ such that $\Nn_k$ approximate the
execution time as closely as $\Nn$.
The value $k$ characterizes the number of observational classes over
the secret inputs in the program and
corresponding network $\Nn_\alpha$ characterizes the secret elements
in each class of observation.

We use an MILP encoding, similar to~\cite{Fischetti2018,narodytska2018verifying,dutta2018output} but in backward
analysis fashions, to count the number of
secret elements in each observational class as characterized by the network
$\Nn_\alpha$.
These counts can then be used to provide a quantitative measure of information
leaks in the program due to the execution times.
Since the function $\beta$ is not directly useful in quantification process,
we use a simpler network model, in our experiments, to compute the reducer
function $\alpha$ as shown in Figure~\ref{fig:arch}.

\section{Experiments}
\subsection{Implementations}
\noindent\textbf{Environment Setup}.
All timing measurements from programs are conducted on an
NUC5i5RYH machine. We run each experiment multiple times and
use the mean of running time for the rest of analysis.
We use a super-computing machine
for the training and analysis of the neural network.
The machine has a Linux Red Hat 7
OS with 24 cores of 2.5 GHz CPU each with 4.8 GB RAM.

\noindent\textbf{Neural Network Learning}.
The neural network model is implemented using TensorFlow~\cite{abadi2016tensorflow}.
We randomly choose 10\% of the data for testing, and the rest for the training.
We use ReLU units as activation functions
and apply mini-batch SGD with the Adam optimizer~\cite{kingma2014adam}
where the learning rate varies from 0.01 to 0.001 for different benchmarks.
For the reducer function of our NN model,
we binarize the output of every layer using the ``straight-through''
technique~\cite{courbariaux2016binarized} to estimate the activation
function in the backward propagation of errors known as
backpropagation~\cite{lecun2015deep}.

\noindent\textbf{Quantification of Information Leaks}.
After training, we analyze the reducer function using mixed integer
linear programming (MILP)~\cite{Fischetti2018,narodytska2018verifying}.
We encode the MILP model in Gurobi~\cite{gurobi} and use the {\em PoolSolutions}
option to retrieve feasible solutions (up to 2 Billions).
For each class of observation (each distinct output of interface layer),
the Gurobi calculates possible solutions from
the secret inputs such that the output value of interface
layer is feasible from those inputs.
We use the number of solutions for each class and apply
Shannon entropy to quantify the amount of information leaks.

\subsection{Micro-benchmarks}
First, we show our approach for finding the (optimal)
number of interface neurons ($k$) from the reducer function.
Then, we show the scalability and usefulness of our approach.
\textit{scalability:} We use the size of
neural network, computation time for learning,
and the computation time for analyzing.
\textit{usefulness:} We consider
the number of classes of observations, the fitness of predictions,
and entropy measures. We also compare the entropy values to ground
truth.

\noindent\textbf{Programs}.
We use two sets of micro-benchmark programs for our studies.
The first one, taken from~\cite{aaai18},
uses the names R\_n where n is the
number of secret bits in the program.
The benchmarks were constructed to exhibit
complex relationships between secret bits that influence the running time.
Each relationship is a boolean formula over the secret input where
the true evaluation triggers a (linear) loop statement over the public inputs.

For Branch\_Loop (B\_L) applications~\cite{FuncSideChan18},
the program does different
computations with different complexities depending on the values
of the secret input. There are four loop complexities:
O($\log(N)$), O($N$), O($N.\log(N)$), and O($N^2$) where $N$
is the public input. Each micro-benchmark B\_L\_i
has all four loop complexities,
and there are $i$ types of each complexity with different constant
factors such as O($\log(N)$) and O($2.\log(N)$) for B\_L\_2.

\begin{figure}[t!]
	\centering
	\includegraphics[width=0.45\textwidth]{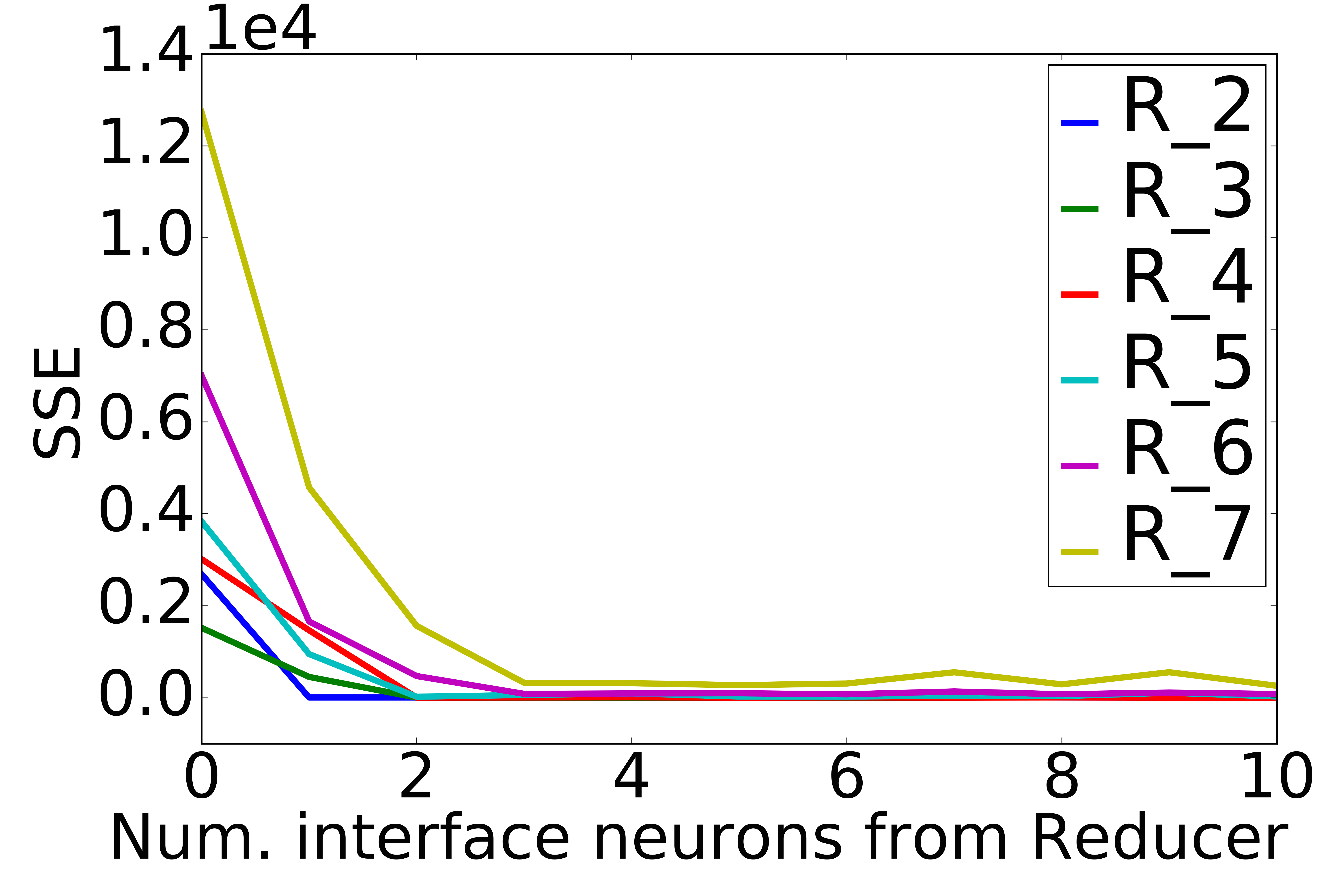}
	\includegraphics[width=0.45\textwidth]{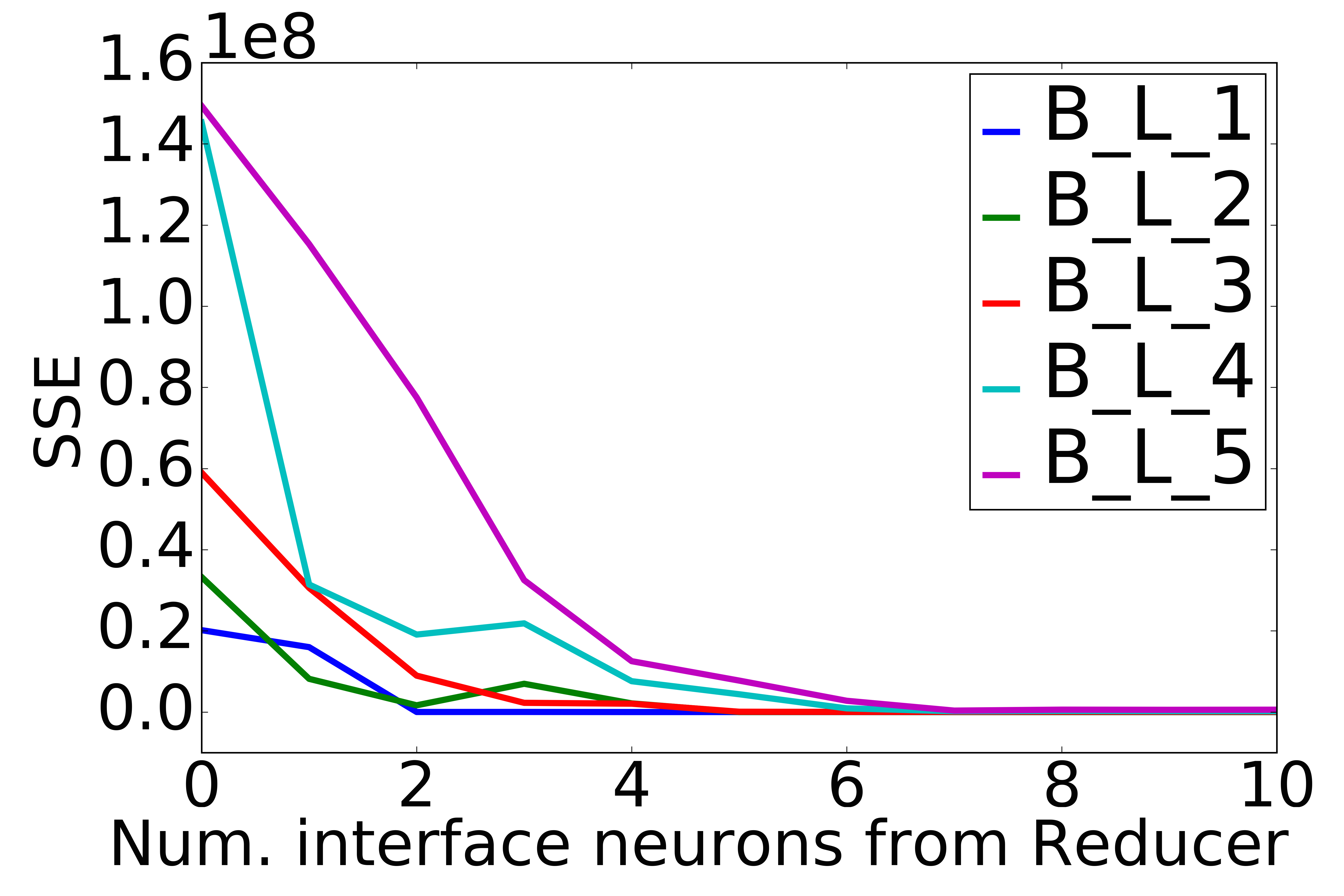}
	\caption{
		(a) The SSE versus the number of interface neurons for R\_n.
		(b) The SSE versus the number of interface neurons for B\_L\_n.
	}
	\label{fig:sse-vs-k-Rn}
	\label{fig:sse-vs-k-BL}
\end{figure}

\noindent\textbf{Optimal number of reducer outputs}.
Since the number of observational classes of
the secret inputs depends on the number of interface neurons ($k$)
from the reducer function, we choose the optimal value for k.
We consider the sum of squared error (SSE)
versus the number of interface neurons (k). We choose a value k
such that the SSE error decreases from 0 to k ($k\geq0$)
and stays almost same for larger values of k.
Figure~\ref{fig:sse-vs-k-Rn} (a) and Figure~\ref{fig:sse-vs-k-BL} (b)
show the plot of the SSE error
vs number of interface neurons for R\_n and B\_L\_n, respectively.
For example, in B\_L\_5, the optimal number of interface neurons
is 7.

\noindent\textbf{Scalability Results}.
Table~\ref{tab:benchmark} shows that our approach is scalable
for learning {\em timing models} of programs.
For example, we could learn the time model of B\_L\_5 program
with the NN model of 7 internal layers and 717 neurons in 25 mins.
In addition, the growth in computation times of learning
is proportional to the growth in the size of networks.
The results show that our approach is scalable
for {\em analyzing} the reducer function of NN.
For this analysis, we only consider the secret parts of NN (shown
with $L_S$ in Table~\ref{tab:benchmark}).
The computation time for the analysis depends on the size
of secret inputs and the size of reducer function.
In B\_L\_5 program with 7 interface neurons
of the reducer, we calculate feasible solutions
over secret inputs for each $2^7$ possible classes.
It takes about 8 minutes to analyze the reducer function
of B\_L\_5 program.
The growth in the computation times of network analysis is
also proportional to the growth in the size of secret inputs and
the size of reducer.
For example, the computation time for the analysis of B\_L\_4
example is increased by almost 12 times in comparison to B\_L\_3,
but the size of input, the interface, and the internal neurons
have increased by two times (from $2^{11}$ to $2^{12}$), two times
(from $2^6$ to $2^7$), and 6 times ($2.5 \times 2.5$), respectively.

\noindent\textbf{Usefulness results}.
We use the statistical metric, coefficient of
determination ($R^2$)~\cite{nagelkerke1991note},
as the fitness indicator of our predictions.
In all benchmarks, $R^2$ is 0.99.
The analysis of the reducer function provides us: 1) whether
a class of observation (a specific value of the reducer output)
is reachable from at least one secret value; 2) how many
secret elements exist in each class of observation.
In B\_L\_5, there are 128 possible values for
the 7 interface neurons. The analysis of network shows that
only 50 values out of 128 are valid and reachable from secret inputs.
To count the number of solutions for each class,
we bound the number of possible solutions to be at most 100.
We use Shanon entropy to measure the amount of information leaks in bits.
For example, in R\_7, the initial Shannon entropy
is $SE_I = 7 (bits)$. We obtain 5 feasible classes:
\{68,16,27,1,16\}. Therefore, after the timing observations,
the conditional Shannon entropy is $SE_O = 5.2 (bits)$.
The amount of information leaks is $SH_L = 1.8 (bits)$.
We note that the initial Shannon entropy may depend
on the number of feasible classes of observations and the
bounds on the possible solutions (see B\_L\_5 as an example).
The ground truth of conditional Shannon entropy is the following:
R\_2=1.19, R\_3=1.44, R\_4=2.42, R\_5=3.42, R\_6=4.0, R\_7=5.0,
and B\_L\_1, B\_L\_2, B\_L\_3, B\_L\_4, B\_L\_5 are all equal to 6.64.

\begin{table*}[!t]
		\caption{
			\#\textbf{R:} number of data records,
			\#\textbf{S:} number of secret bits,
			\#\textbf{P:} number of public bits,
			\textbf{$L_S$:} the size of secret part (reducer function) of NN,
			\textbf{$L_P$:} the size of public part of NN,
			\textbf{$L_J$:} the size of joint part of NN,
			\textbf{$\alpha$:} learning rate,
			\textbf{$R^2:$} coefficient of determination,
			\textbf{$T_L:$} the computation time (s) for learning NNs,
			\textbf{$T_A:$} the computation time (s) for analyzing reducer functions,
			\#\textbf{k}: number of interface neurons in the reducer function,
			\#\textbf{K}: number of (feasible) classes of observations,
			\textbf{$SE_I$}: initial Shannon entropy (before any observations),
			\textbf{$SE_O$}: remaining Shannon entropy after timing observations.
		}
		\label{tab:benchmark}
		 \resizebox{\textwidth}{!}{
		\begin{tabular}{ || l | c | c | c | c | c | c | c | c | c | c | c | c | c | c || }
			\hline
			&      &    &    &    &    &   &   &   &  &   &   &  &  & \\
			App(s) & \#\textbf{R} & \#\textbf{S} & \#\textbf{P}  & \textbf{$L_S$} & \textbf{$L_P$} &
			\textbf{$L_J$} & \textbf{$\alpha$} & \textbf{$R^2$} & \textbf{$T_L$} &
			\textbf{$T_A$} & \#\textbf{k} & \#\textbf{K} & \textbf{$SE_I$} & \textbf{$SE_O$}  \\ \hline
			R\_2 & 400 & 2 & 7 & $[5 \times 1]$  & $[5]$ & $[10]$  & 1e-2  & 0.99 & 91.7  & 0.1  & 1 & 2 &  2.0 & 1.19   \\ \hline
			R\_3 & 800 & 3 & 7  & $[10 \times 2]$  & $[10]$ & $[20]$ & 1e-2 & 0.99  & 91.7  & 0.1 & 2 & 3  & 3.0 & 1.44   \\ \hline
			R\_4 & 1,600 & 4 & 7 & $[10 \times 2]$ & $[10]$ & $[20]$ & 1e-2  & 0.99  & 90.3  & 0.1  & 2 & 4  & 4.0  &  2.32 \\ \hline
			R\_5 & 3,200 & 5 & 7  & $[10 \times 10 \times 2]$  & $[10]$ & $[20]$ & 1e-2  & 0.99  & 127.3  & 0.2 & 2 & 3  & 5.0  & 3.4  \\ \hline
			R\_6 & 6,400 & 6 & 7  & $[10 \times 10 \times 3]$  & $[10 \times 10]$ & $[20]$ & 1e-2  & 0.99  & 168.4 & 0.5  & 3 & 5  & 6.0  & 4.0 \\ \hline
			R\_7 & 12,800 & 7 & 7 & $[20 \times 20 \times 3]$  & $[10 \times 10]$ & $[20]$ & 1e-2  & 0.99  & 185.4  & 1.8  & 3 & 5  & 7.0  & 5.0 \\ \hline
			B\_L\_1 & 756  &  9 & 7 & $[20 \times 20 \times 4]$  & $[10]$ & $[20 \times 20]$ & 1e-2  & 0.99 & 124.3 & 0.3 & 4 & 5 & 8.97  & 6.43 \\ \hline
			B\_L\_2 & 1,512  & 10 & 7 & $[40 \times 5]$  & $[10]$ & $[40 \times 40]$ & 1e-2 & 0.99 & 129.4 & 2.5  & 5 & 10  & 9.97  & 6.40   \\ \hline
			B\_L\_3 & 3,024  & 11 & 7 & $[20 \times 20 \times 6]$  & $[10]$ & $[100 \times 100]$ & 5e-3 & 0.99 & 346.0 & 18.6 & 6 & 16 & 10.64 & 6.39 \\ \hline
			B\_L\_4 & 6,048  & 12 & 7 &  $[50 \times 50 \times 7]$ & $[10]$ & $[200 \times 200]$ & 5e-3 & 0.99 & 889.8 & 216.7 & 7 & 39 & 11.93 & 6.0 \\ \hline
			B\_L\_5 & 12,096  & 13 & 7 & $[50 \times 50 \times 7]$ & $[10]$ & $[200 \times 400]$ & 2e-3 & 0.99 & 1,411.0 & 496.6  & 7 & 50 & 12.29 & 6.1 \\ \hline
		\end{tabular}
		}
\end{table*}

\label{sec:experiment}

\section{Case Studies}
\begin{figure}[t!]
	\centering
	\includegraphics[width=0.5\textwidth]{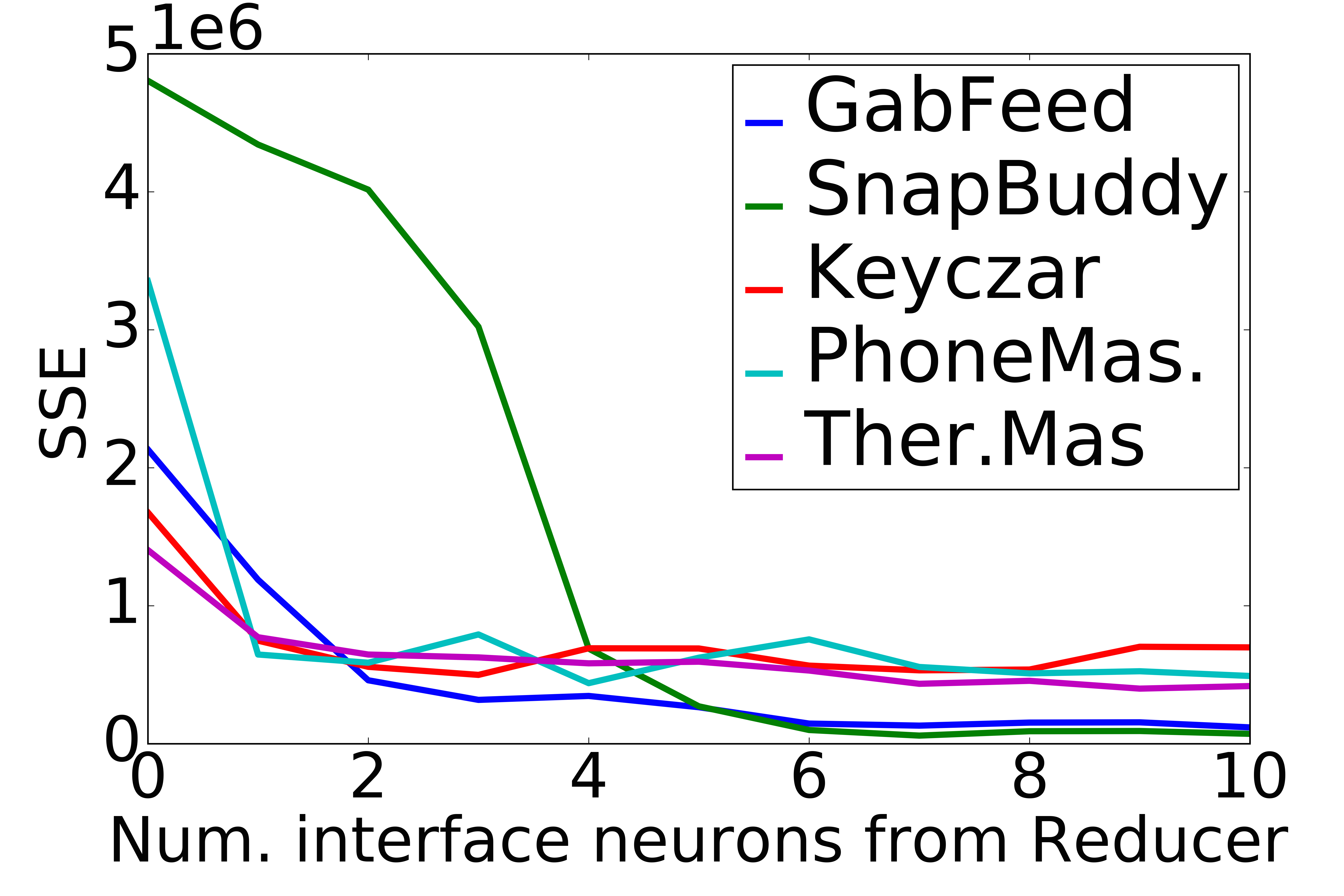}
	\caption{
		The SSE versus the number of interface neurons k for case-study
		application.
	}
	\label{fig:sse-vs-k-2}
\end{figure}

Table~\ref{tab:case-study} summarizes 5 real-world Java applications used as
case studies in this paper. Table~\ref{tab:case-study} has similar structure to
Table~\ref{tab:benchmark} in Section~\ref{sec:experiment} and also
lists the number of methods in the applications.
Figure~\ref{fig:sse-vs-k-2} shows the SSE (error) vs the number of
interface neurons of the reducer function for case-study applications.
The main research questions are
``Does our approach of using neural networks for side-channel analysis
of real-world applications 1) scale well, 2) learn timing models
accurately, and 3) give useful information about the strength of leaks?''

\subsection{GabFeed}
Gabfeed~\cite{sources} is a Java web application with 573
methods implementing a chat server
~\cite{DBLP:conf/ccs/ChenFD17}.
The application and its users can mutually authenticate
each other using public-key infrastructure.
The server takes users' public key and its own private key
and calculate a common key.

\noindent\textit{Inputs}. We consider the secret and public keys with
1,024 bits. We generate 65,908 keys (combination of secret and public keys)
that are uniformly taken from the space of secret and public inputs.

\noindent\textit{Neural Network Learning}.
We learn the timing model of GabFeed for generating
common keys with $R^2=0.952$ where we set the
learning rate to 0.01.
The NN model consists of 1,024 binary secret
and 1,024 binary public inputs. The network has
more than 600 neurons. The optimal number of
neurons for interface layer is 6.
It takes 40 minutes to learn the timing model
of GabFeed application.

\noindent\textit{Security Analysis}.
Since the output of the reducer is 6 bits, there are at
most 64 classes of observations. Our analysis shows that
there are only 26 feasible classes.
With the assumption that each class can have at most 10,000 solutions,
the initial Shannon entropy is 18.0 bits.
By observing the 26 classes through timing side channels,
the remaining Shannon entropy becomes 13.29 bits. Therefore,
the amount of leaks is 4.71 bits. The security analysis
of NN takes 300 minutes.

\noindent\textit{Research Questions}. To answer our research questions:
{\em Scalability:} The neural network model has 606 neurons.
It takes 40 minutes to learn the time model of GabFeed
applications. It takes 300 minutes to analyze the reducer function and obtain
the number of elements in each class.
{\em Usefulness:} We learn the time model of GabFeed as the function of
public and secret inputs with $R^2 = 0.952$. Our analysis shows
that there are 26 classes of observations over the secret inputs, and
4.71 bits of information about the secret key is leaking.

\begin{table*}[!t]
	\caption{Case Studies. Legends similar to
	Table~\ref{tab:benchmark} in Sec.~\ref{sec:experiment}
	except that \#\textbf{M} shows the number of methods
	in the application and $SE_L$ is the difference
	between $SE_O$ and $SE_I$ and shows the amount of information
	leaked in bits.
	}
	\label{tab:case-study}
	\resizebox{\textwidth}{!}{
		\begin{tabular}{ || l | c | c | c | c | c | c | c | c | c | c | c | c | c | c || }
			\hline
			&      &    &    &    &    &   &   &   &  &   &   &   &  & \\
			App(s) & \#\textbf{M} & \#\textbf{R} & \#\textbf{S} & \#\textbf{P}  & \textbf{$L_S$} & \textbf{$L_P$} &
			\textbf{$L_J$} & \textbf{$\alpha$} & \textbf{$R^2$} & \textbf{$T_L$} &
			\textbf{$T_A$} & \#\textbf{k}  & \#\textbf{K} & \textbf{$SE_L$} \\ \hline
			GabF. & 573 & 65,908  & 1,024  & 1,024 & $[50\times50\times6]$  & $[100]$ & $[200\times200]$ & 1e-2 & 0.95 & 2,410 & 18,010 & 6 & 26 & 4.7 \\ \hline
			Snap. & 3,071 & 6,678  & 30 & 4 & $[30\times30\times6]$ & $[10]$ & $[100]$ & 1e-2 &
			0.98 & 579  & 17 & 6 &  8 & 3.0   \\ \hline
			Phon. & 101 & 3,043 & 82 & 11 &  $[50\times50\times10]$ & $[5]$ & $[100]$ & 8e-3 &
			0.99 & 566  & 10,151  & 10 & 60 & 5.9 \\ \hline
			Ther. & 53 & 10,000 & 11 & 4 & $[100\times100\times9]$ & $[100]$ & $[200\times200]$
			& 1e-3 & 0.8 & 4,236 & 5,148 & 9 & 9 & 7.0  \\ \hline
			PassM. & 6  & 211,238 & 14 & 30 & $[50\times50\times3]$ & $[50]$ & $[100]$ &  1e-2  &
			0.98 & 202 & 16  & 3 & 4  & 1.6  \\ \hline
		\end{tabular}
	}
\end{table*}

\subsection{SnapBuddy}
\label{sec:case-SB}
SnapBuddy is a mock social network application where each user
has their own profiles with a photograph~\cite{tizpaz2017discriminating}.
The profile page is publicly accessible.

\noindent\textit{Inputs}. The secret is the identity of a user (among 477
available users in the network) who is currently interacting with the server.
The public is the size of each profile (from 13 KB to 350KB).
Note that the size of profiles are observable from generated network traffics.

\noindent\textit{Neural Network Learning}.
We consider the response time of the SnapBuddy application to download
public profiles of 477 users in the system~\cite{sources}.
We learn the response time using a neural network
with 176 neurons and 6 neurons
in the interface layer.
The accuracy of neural network model in predicting
response times based on the coefficient of determination
is 0.985 where we set the
learning rate to be 0.01. The learning takes less than 10 minutes.

\noindent\textit{Security Analysis}.
Our analysis finds only 8 classes of observations reachable out of 64.
Since the number of users (secrets) in the current database is fixed to 477, we
assume there can be at most 60 users in each class. The initial Shannon
entropy is 8.91 bits. The remaining Shannon entropy after observing the
execution times and obtaining
the classes of observations with their characteristics
is 5.91 bits. The amount of information leaks is 3.0 bits. The analysis
of reducer function takes less than 17 seconds.

\noindent\textit{Research Questions}. To answer our research questions:
{\em Scalability:}
It takes less than 10 minutes to learn the time model of SnapBuddy.
It takes only 16.6 seconds to calculate feasible solutions for all of
feasible classes of observations.
{\em Usefulness:} We learn the time model of SnapBuddy as a function of
public and secret inputs with $R^2 = 0.985$. Our analysis shows
that there are 8 classes of observations over the secret inputs, and
3.0 bits of information about users' identities are leaking.

\subsection{PhoneMaster}
Phonemaster~\cite{sources} is a record keeping service for tracking
phone calls and bills. The identity of a user who submits a request is
secret, while the generated traffic from the interaction is public.

\noindent\textit{Inputs}.
There are at most 150 users. For each user, we send a random
command from the set of possible commands.

\noindent\textit{Neural Network Learning}.
We use a neural network with 215 neurons. We find out
that the optimal number of neurons in the interface layer
is 10. We learn the time model of phoneMaster in
less than 10 minutes with $R^2 = 0.993$.

\noindent\textit{Security Analysis}.
We analyze the reducer function of NN and
find out that 60 classes of observations are feasible.
We assume that there can be at most 3 users in each class.
The initial Shannon entropy is 7.49 bits.
The remaining Shannon entropy after observing the
execution times is 1.58. The amount of information leakage is 5.91 bits.
This shows that almost everything about the identity of users is leaking.
The computation time of analysis is about 169 minutes.

\subsection{Thermomaster}
Thermomaster~\cite{sources} is a temperature control
and prediction system.
The program takes the goal temperature (secret inputs)
and the current temperature (public inputs) to simulate the controller
for matching with the goal temperature.

\noindent\textit{Inputs}.
The goal temperature is between -10,000 and +10,000 and the current
temperature is between -250 and +250. We generate 10,000 inputs
uniformly from the space of goal and current temperatures.

\noindent\textit{Neural Network Learning}.
We use a NN model with 709 neurons. The optimal
number of neurons for the interface layer is 9.
We learn the timing models of thermomaster in
70 minutes with $R^2 = 0.80$.

\noindent\textit{Security Analysis}.
We analyze the reducer function of NN and find out that
only 9 classes of observations are feasible.
The initial Shannon entropy is 11.0 bits.
The remaining Shannon entropy after observations is 4.0.
Therefore, 7.0 bits of information about the goal temperature
are leaking through timing side channels.
The computation time of analysis is less than 86 minutes.

\subsection{Password Matching (Keyczar)}
We consider a vulnerability in a password matching
algorithm similar to the side-channel vulnerability
in Keyczar library~\cite{Keyczar}.
This vulnerability
allows one to recover the secret password through
sequences of oracles where the attacker learns
one letter of the secret password in each step.

\noindent\textit{Inputs}.
The secret input is target password stored in a
server, and the public input is a guess oracle.
We use libFuzzer~\cite{libFuzzer} to generate
21,123 guesses for randomly selected passwords.
We assume a password is at most 6 (lower-case) letters.

\noindent\textit{Neural Network Learning}.
We use a neural network with 253 neurons. The optimal
number of neurons for the interface layer is 3.
We learn the time model of the password matching algorithm in
202.3 seconds with $R^2{=}0.976$.

\noindent\textit{Security Analysis}.
There can be at most 8 classes of observations.
Our analysis shows
only 4 classes of observations are feasible.
The initial Shannon entropy is 14 bits.
The 4 classes (obtained from timing observation)
have the following number of elements: $\{48, 6760, 4309, 5269\}$.
So, the remaining entropy (after observing the classes through the time
model) is 12.4 bits. It takes about 16 seconds to
analyze the secret parts of NN and quantify the information leaks.

\label{sec:case-study}

\section{Related Work}
\noindent\textbf{Modeling Program Execution Times.}
Various techniques have been applied to model and predict computational
complexity of software
systems~\cite{goldsmith2007measuring,aaai18,arar2015software}. Both
\cite{goldsmith2007measuring} and \cite{aaai18} consider cost measures such as
execution time and predict the cost as a function of input features such as the
number of bytes in an input file. The works~\cite{goldsmith2007measuring,aaai18} are
restricted to certain classes of functions such as linear functions,
while the neural network techniques can model
arbitrary functions. Additionally, both techniques require feature engineering:
the user needs to specify some features such as size or work-load features.
However, neural network models do not
require this and can automatically discover important features.

\noindent\textbf{Neural Networks for Security Analysis.}
Neural network models have been used for software security analysis. For
example, the approach in~\cite{tang2016deep} uses the deep neural network for
anomaly detection in software defined networking (SDN). The
framework~\cite{li2018vuldeepecker} uses a deep neural network model for detecting
vulnerabilities such as buffer and resource management errors. We use
neural network models to detect and quantify information leaks
through timing side channels.

\noindent\textbf{Dynamic Analysis for Side-Channel Detections.}
Dynamic analysis has been used for side-channel detections
~\cite{milushev2012noninterference,DBLP:conf/icse/nilizadeh,profit2019}.
Diffuzz~\cite{DBLP:conf/icse/nilizadeh} is a fuzzing techniques for
finding side channels. The approach
extends AFL~\cite{AFL} and KELINCI~\cite{kersten2017poster}
fuzzers to detect side channels. The goal of Diffuzz is
to maximize the following objective: ${\delta = |c(p,s_1) - c(p,s_2)|}$,
that is, to find two distinct secret values $s_1,s_2$ and a public value $p$
that give the maximum cost ($c$) difference.
The work~\cite{DBLP:conf/icse/nilizadeh}
uses the noninterference notion of side channel leaks. Therefore,
they do not quantify the amounts of information leaks.
The cost function
in~\cite{DBLP:conf/icse/nilizadeh} is the number of byte-code executed,
whereas we consider the actual execution time in a fixed environment.
Note that our approach can be used with the abstract cost model such as
the byte-code executed in a straight-forward fashion.
Diffuzz~\cite{DBLP:conf/icse/nilizadeh}
can be combined with our technique to generate inputs
and quantify leaks.

\noindent\textbf{Static Analysis for Side-Channel Detections.}
Noninterference was first introduced by
Goguen and Meseguer~\cite{goguen1982security}
and has been widely used to enforce confidentiality properties in various
systems~\cite{sabelfeld2003language,terauchi2005secure,almeida2016verifying}.
Various
works~\cite{DBLP:conf/ccs/ChenFD17,antonopoulos2017decomposition}
use static analysis
for side-channel detections based on noninterference notion.
The work~\cite{DBLP:conf/ccs/ChenFD17} defines
$\varepsilon$ bounded noninterference that requires the resource usage behavior
of the program executed from the same public inputs differ at most $\varepsilon$.
Chen et al.~\cite{DBLP:conf/ccs/ChenFD17}
use Hoare Logic~\cite{barthe2004secure} equipped with taint
analysis~\cite{livshits2005finding} to detect side channels.
These static techniques including~\cite{DBLP:conf/ccs/ChenFD17}
rely on the taint analysis that is computationally difficult
for real-world Java applications.
The work~\cite{landman2017challenges} reported that 78\% of 461 open-source
Java projects use dynamic features such as
reflections that are problematic for static analysis.
In contrast, we use dynamic analysis that handles the reflections
and scales well for the real-world applications.
In addition, Chen et al.~\cite{DBLP:conf/ccs/ChenFD17}
answer either `yes' or `no' to the existence of side channels, which
is restricted for many real-world applications that may need to disclose
a small amounts of information about the secret. However,
our approach quantifies the leaks using entropy measures.

\noindent\textbf{Quantification of Information Leaks.}
Quantitative information
flow~\cite{backes2009automatic,smith2009foundations,KB07} has been used for
measuring the strength of side channels. The work~\cite{backes2009automatic}
presents an approach based on finding the equivalence relation over secret
inputs. The authors cast the problem of finding the equivalence relation as a
reachability problem and use model counting to quantify information leaks. Their
approach works only for a small program, limited to a few lines of code, while
our approach can work for large applications. In addition, they consider
the leaks through direct observations such as program outputs or public
input values. In contrast, we consider the leaks through timing side channels,
which are non-functional aspects of programs.
Sidebuster~\cite{zhang2010sidebuster} combines static and dynamic
analyses for detection and quantification of information leaks.
Sidebuster~\cite{zhang2010sidebuster} also relies on taint analysis
to identify the source of vulnerability. Once the source identified,
Sidebuster uses dynamic analysis and measures the amounts of information leaks.
The information leaks in Sidebuster~\cite{zhang2010sidebuster} is
because of generated network packets, while our information leaks
are through timing side channels.

\noindent\textbf{Hardening Against Side Channels.}
Hardening against side channels can be broadly divided
to mitigation and elimination approaches.
The mitigation approaches~\cite{kopf2009provably,askarov2010predictive,DBLP:conf/cav/Tizpaz-NiariC019}
aim to minimize the amounts of information leaks, while considering
the performance of systems. The goal of elimination
approaches~\cite{agat2000transforming,wu2018eliminating,eldib2014synthesis}
is to completely transform out information leaks without considering
the performance burdens. Our techniques
can be combined with the hardening methods to mitigate
or eliminate information leaks.

\noindent\textbf{Other Types of Side Channels.}
Sensitive information can be leaked through other side channels
such as power consumptions~\cite{eldib2014formal,wang2019mitigating},
network traffics~\cite{CWWZ10}, and
cache behaviors~\cite{guo2018adversarial,Sung:2018,doychev2015cacheaudit,wang2017cached}.
We believe our approach could be useful for these types of
side channels, however, we left further analysis for future work.

\label{sec:related}

\section{Conclusion and Discussion}
We presented a data-driven dynamic analysis for detection and
quantifying information leaks due to execution times of programs.
The analysis performed over a specialized NN architecture
in two steps: first, we utilized neural
network objects to learn timing models of programs and second,
we analyzed the parts of NNs related to secret inputs to detect
and quantify information leaks.
Our experiences showed that NNs learn timing
models of real-world applications precisely. In addition,
they enabled us to quantify information leaks, thanks
to the simplicity of NN models in comparison to program models.

Throughout this work,
we assume that the analyzer would be able to construct interesting
inputs either with fuzzing tools, previously reported bugs,
or domain knowledges.
Nevertheless, we demonstrate practical solutions to
generate inputs in each example with emphasis on the recent development
in fuzzing for side-channel analysis~\cite{DBLP:conf/icse/nilizadeh}.

Furthermore, our dynamic analysis approach can not
prove the absent of side channels. Our NN model learns and generalizes
the timing models for the observed program behaviors and is limited to
observed paths in the program.
We emphasize that
the proof is also difficult for static analysis.
Although static analysis can prove the absent of
bugs or vulnerabilities in principle, the presence
of dynamic features such as reflections in Java applications
is problematic and can cause false negative in
static analysis (see Limitations Section
in~\cite{DBLP:conf/ccs/ChenFD17}).

For {\em future work}, there are few interesting
directions. One idea is to develop a SAT-based algorithm,
similar to DPLL, on top of MILP algorithms to calculate
the number of solutions more efficiently.
Another idea is to define
threat models based on the attackers capabilities to
utilize neural networks for guessing
secrets.

\label{sec:discussion}

\paragraph{Acknowledgements.}
The first author thanks Shiva Darian for proofreading and providing
useful suggestions.
This research was supported by DARPA under agreement FA8750-15-2-0096.

\clearpage

\bibliographystyle{splncs04}
\bibliography{papers}

\end{document}